\documentclass[12pt,a4paper]{article}

\usepackage{jheppub}

\usepackage[utf8]{inputenc}
\usepackage{graphics}     
\usepackage{graphicx}
\usepackage{caption}
\usepackage{url}          
\usepackage{bm,bbm}          
\usepackage{mathtools,amsmath,amsthm,amsfonts,amssymb,lipsum,xcolor}
\usepackage{enumitem}
\usepackage{hyperref}
\usepackage{xcolor}
\usepackage{ulem}
\usepackage{tikz}
\usetikzlibrary{positioning,calc,intersections}
\usepackage{cancel}

\newcommand{\comm}[2]{\left[#1,#2\right]}

\newcommand{\rn}[1]{\mathbb{R}^{#1}}
\newcommand{\de}[1]{\partial_{#1}}

\newcommand{\del}[2]{\frac{\delta #1}{\delta #2}}

\newcommand{\matrice}[3]{#1_{#2}^{\,\,\,#3}}
\newcommand{\Matrice}[3]{#1^{#2}_{\,\,\,#3}}

\newcommand{\labeledset}[3]{\left\{#1_{#2}\right\}_{#3}}
\newcommand{\Labeledset}[3]{\left\{#1^{#2}\right\}_{#3}}

\newcommand{\cn}[1]{\mathbb{C}^{n}}

\newcommand{\inversem}[1]{\frac{g_{#1}}{g^{00}}}

\title{Pseudo-orthogonal Yang-Mills theories and connections to gravity}

\author{Giovanni Mistretta,}
\emailAdd{giovannimistretta00@gmail.com}
\author{Tomislav Prokopec}
\emailAdd{t.prokopec@uu.nl}

\affiliation{Institute for Theoretical Physics, Spinoza Institute \& EMME$\Phi$,
	\\
	Utrecht University, Princetonplein 5,
	\\
	3584 CC Utrecht, The Netherlands}

\abstract{
\date{today}\\
We formulate gauge theories on noncompact Lorentzian manifolds.
For definiteness we choose 
an SO(1,4) gauge theory -- the isometry group of the five dimensional Minkowski space.  We make use of the natural inner product to
construct the Yang-Mills gauge action on four dimensional spacetime, on which the natural 
tetrad and metric are induced, thus breaking the symmetry to that of general relativity. 
In the low energy limit -- if a suitable gauge field condensate  develops -- the theory reduces to the Cartan-Einstein gravity, which harbors nondynamical torsion, 
and is consistent with all observations. 
We also discuss how to couple our gauge theory of gravity to scalar and vector matter.
The Hamiltonian analysis shows that the theory possesses no Ostrogradsky instabilities, however
it harbors a kinetic instability. We conjecture that 
such a kinetic instability can be removed either by generalizing the theory to the nonlinear 
Born-Infeld theory, or by constraining the kinetic instability. This work is an attempt to 
formulate gravity as a unitary, renormalizable gauge theory without instabilities, in which
the fundamental propagating degrees of freedom are in the spin-one tetrad connection.

}

\begin{document}

\maketitle

\section{Introduction}
The first gravitational theory oversaw the birth of theoretical physics, perhaps the last one will see its end.
For almost a century, researchers from all over the world tried to properly quantize the gravitational field.
Up to now, no convincing solution to this problem has been found.
Starting from general relativity in 1915, the theory has been extended and generalized in many different ways (see~\cite{Buchbinder:2021wzv}
 for a review).
Among these theories, string theory is the only one in which Einstein's field equations are obtained without adding by hand the Hilbert--Einstein action 
to the theory, or by introducing it as a counterterm as in some induced gravity theories. 
Indeed, gravity  (on a world-sheet) emerges in string theory by setting to zero the beta functions of the theory, imposing conformality at the quantum level~\cite{Tong:2009np}.
The goal of this paper is to show that there exists another class of theories obtained by a suitable generalization of the well-known Yang-Mills theories
that contains Einstein theory in its torsionless low energy limit.

As the Standard Model teaches us, fundamental interactions in Nature are mediated (at low-energy scales) by gauge fields of the Yang-Mills type.
These theories are proved to be renormalizable, which motivates us to look
at gauge theories for a possible solution to the renormalizability problem of quantum gravity.
Moreover, it is well-known that the renormalization of flat spacetime theories generically induces higher order geometric scalars with respect to the Ricci scalar present in the Hilbert action~\cite{tHooft:1974toh}.
Usually these terms provide instabilities of the Ostrogradsky kind~\cite{Woodard:2015zca}
, due to the presence in the action of quadratic second-order time derivatives with respect to some components of the metric tensor.
Since the Yang-Mills theory fundamentally provides a theory of a field strength squared, it seems natural to use an appropriate gauge group in order to reproduce at least these counterterms in a stable way.
This is possible due to the fact that a Yang-Mills theory is a first order formalism similar to the Palatini formalism in GR ({\it i.e.} the variational principle of Hilbert action for which Christoffel symbols are considered free and they are not the Levi-Civita connection~\cite{Misner:1973prb}).
Another reason why Yang-Mills theories have a chance of explaining gravity is the {\it geometrical structure} that 
underlies them.
General relativity and Yang-Mills theories~\cite{Hamilton:2017gbn} are two of the most important examples of differential geometry in theoretical physics.
It is therefore natural to pose
the question whether one can use the geometry of gauge theories 
to derive general relativity. 

Guided by the fact that linear gravity is a spin-2 field theory~\cite{Fierz:1939ix}, 
and by the observation that general relativity (GR) can be viewed as the low-energy limit
of the more general effective theory of gravity~\cite{Donoghue:2012zc}, 
we are inspired to use spin-1 Yang-Mills fields (whose product is known to form a 
spin-2 representation) 
to describe a theory that reduces to GR in the low-energy limit.~\footnote{Interestingly, Witten proved~\cite{Witten:1988hc}
that the 2+1 general relativity is equivalent to the Chern-Simons theory -- a topological Yang-Mills theory.} 
This work is an attempt to define a {\it geometrical Yang-Mills theory} in four spacetime dimensions, whose
low energy limit is GR.

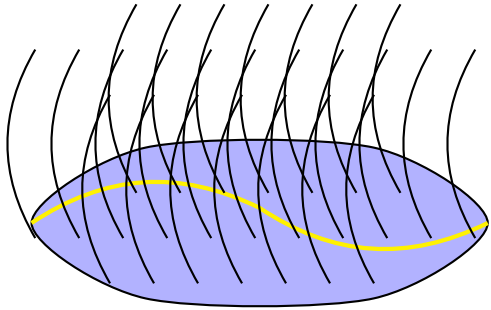
\begin{figure}
    \centering
    \begin{tikzpicture}
        \draw[thick, fill=blue!30] plot [smooth cycle] coordinates {(1,0) (2.5,1) (5.5,1) (7,0) (5.5,-1) (2.5, -1)};
        \node (M) at (4, 0.5) {};
        \draw[yellow, ultra thick, name path=path] (1,0) to[bend left] (4,0.2) to[bend right] (7,0);       
        \foreach \x in {0.1, 1.3}
        {
        \foreach \y in {-6+\x, -4+\x, -2+\x, 0+\x, 2+\x, 4+\x, 6+\x}
        {
            \coordinate (Start) at ($(M.east)!\y!(M.west)+(0,-\x)$);
            \coordinate (End) at ($(Start) + (0,2.5)$);
            \draw[-, thick] (Start) to[bend left] (End);        
         }
        }
        \foreach \x in {0.7}
        {
            \foreach \y in {-10+\x, -8+\x, -6+\x, -4+\x, -2+\x, 0+\x, 2+\x, 4+\x, 6+\x, 8+\x, 10+\x}
        {
            \coordinate (Start) at ($(M.east)!\y!(M.west)+(0,-\x)$);
           
            \coordinate (End) at ($(Start) + (0,2.5)$);
            \draw[-, thick] (Start) to[bend left] (End);
        }
        }
    \end{tikzpicture}
   \caption{Spacetime (yellow) imbedded in de Sitter space (purple); the fibers represent the frame bundle of the de Sitter group.}
\label{figure one}
\end{figure}
Following the work by James T. Wheeler~\cite{Wheeler:2013ora} and Juan Trujillo~\cite{Trujillo:2013}, 
we studied the possibility of obtaining a gravitational theory such as Weyl squared gravity from a Yang-Mills theory of the conformal group.
Soon we realized that, in order to twist the geometry of spacetime with the geometrical structure of their gauge theory, we needed a way to define the metric tensor in a non-trivial gauge theoretical way.
Here we develop a new class of Yang-Mills theories, called geometrical Yang-Mills theory.
We will show that it is possible to define a metric for spacetime using part of the gauge fields as cotetrad fields.
Pseudo-orthogonal groups will result as the best choice for our gauge groups.
In particular, de Sitter theory SO(1,4) will reduce, in its torsionless low energy limit, to general relativity with the appearance of a gauge theoretical Planck mass and cosmological constant.
Other groups are also worth considering, such as ISO(1,3) and SO(2,3),
which are the isometry groups of four-dimensional Minkowski and anti-de Sitter space, respectively.

Earlier attempts to formulate gravity as a gauge theory
include the work of MacDowell and Mansouri~\cite{MacDowell:1977jt}, in which 
a geometric gauge theory of gravity was constructed.
{The authors used  the (incomplete) Levi-Civita tensor of the 
symmetry space to define the inner product  and that the tetrad field vanishes on-shell. The choice of the inner product was declared natural, 
but never properly justified. However, due to the fact that 
their inner product is 
{\it not} compatible with the gauge covariant derivative, the theory is {\it not topological} and --
in its low-energy limit -- it reduces to general relativity (in vacuum).
 Wilczek~\cite{Wilczek:1998ea,Addazi:2020cax} modified the theory by
introducing a scalar field in the adjoint representation 
and a symmetry-breaking potential, thus making the theory manifestly 
gauge invariant, but still with an inner product  non-compatible with the gauge covariant derivative.
BF theories can be also used to unify matter and gravity. In Ref.~\cite{Alexander:2012ge} 
gravity is described by an $SU(2)_R$ connection, which 
at high energies gets unified with the $SU(2)_L$ of the electroweak theory.

In this work we revisit the question of 
the natural inner product on the group space, and opt for the one constructed from the Killing metric on the group space,
which adorns the inner product of the Standard Model gauge theories,
and which {\it is} compatible with the covariant derivative.
In our rendering of the theory the spacetime action emerges as a projection of the 5-dimensional space (manifold) on which the de Sitter group fibration is defined, as illustrated in figure~\ref{figure one}. This projection (in yellow) then selects 
a natural tetrad, with respect to which the spacetime volume is defined, thus breaking the original gauge symmetry
down to SO(1,3) -- the symmetry of general relativity.

The two above mentioned theories -- the MacDowell-Mansouri and Wilczek theories  -- can be considered as special realizations of the more general BF-theories~\cite{Freidel:1999rr} (or slightly more general Holst theories~\cite{Holst:1995pc}), reviewed in~\cite{Celada:2016jdt}, 
with a particular choice of the B-tensor.
These theories are topological or not, depending on whether the B-tensor is chosen to be non-dynamical
or dynamical, respectively. Various versions of the BF-theory have been studied in the literature~\cite{Baez:1999sr,Mikovic:2006yp,Durka:2010zx,Alexander:2011jf}, see~\cite{Celada:2016jdt} for a more complete account of the literature.

The paper is organised as follows. 
Section \ref{sec:first}, is dedicated to the definition of geometrical Yang-Mills theory showing the necessity of studying pseudo-orthogonal gauge groups.
In Section \ref{sec:second} we develop the de Sitter gauge theory we have already mentioned.
It is the easiest consistent example of geometrical Yang-Mills theory that contains the Ricci scalar as part of the action.
Finally, in Section \ref{sec:third} we introduce a Hamiltonian formalisms for generic Yang-Mills theories in curved dynamical spacetimes.
In particular we focus on the constraints of the theory establishing their class and their self-consistency conditions.
We conclude by addressing the missing steps of a proper instability analysis, and by giving 
outlooks for the theories we have developed.


\section{Geometrical Yang-Mills theories}
\label{Geometrical Yang-Mills theories}
\label{sec:first}

This section is dedicated to the mathematical construction of some special Yang-Mills theories that we call {\it geometrical}.
The reason for this name is that we will study the implications of using part of the gauge connection as a tetrad.
Throughout the following fix a manifold $M$ and a principal $G$-bundle $P\rightarrow M$ ($G$ being a Lie group).
Our goal is to define the metric on $M$ through the gauge connections.
Let's consider the gauge connection to be $\pmb{\omega}=\tilde{\pmb{\rho}}+\pmb{\sigma}$, where $\pmb{\omega},\tilde{\pmb{\rho}},\pmb{\sigma} \in T^{*}P$, where $T^{*}P$ denotes the contangent space.
We consider $\tilde{\pmb{\rho}}$ to be the part of the gauge connection that defines the metric, while the latter is independent from $\pmb{\sigma}$.
The distinction in $\tilde{\pmb{\rho}}$ and $\pmb{\sigma}$ generates a distinction also in the gauge algebra.
This happens since this forms take value in the Lie algebra $\mathfrak{g}$ of the gauge group.
Next, we introduce $\labeledset{a}{i}{i=1,...,N_{A}}$ and $\labeledset{b}{j}{j=1,...,N_{B}}$ with $N_{A}+N_{B}=N=\text{dim}(\mathfrak{g})$ such that we can write,
\begin{equation}
    \label{eq:connectionsplitting}
    \pmb{\omega}=\omega^{i}\otimes \hat e_{i}=\tilde{\pmb{\rho}}+\pmb{\sigma}=\tilde{\rho}^{i}\otimes a_{i}+\sigma^{j}\otimes b_{j},
\end{equation}
where $\labeledset{\hat e}{i}{i=1,...,N}$ is a basis for the Lie algebra $\mathfrak{g}$. 
We are then tempted to define the metric tensor as,
\begin{equation}
    \label{eq:metricprotodefinition}
    g \equiv \eta_{ab}\tilde{\rho}^{a}\otimes\tilde{\rho}^{b},
\end{equation}
where $\eta_{ab}\, (a,b=0,1,2,3)$
is the Minkowski metric, {\it i.e.} $\text{diag}(-1,1,1,1)$.
Notice that, if we intend to consider the fields $\tilde{\rho}^{i}$ equivalent to standard tetrad fields, we need more properties.
In particular, $N_{A}$ must be equal to $n$, dimensionality of $M$.
Moreover, dimensional analysis gives a dimensionless connection 1-form (for the non-tetrad fields) and the dimension of an inverse mass for the tetrad 1-form.
Since the connection 1-forms and the tetrad fields are part of the same connection 1-form they need to have the same mass dimension.
We can then write $\rho^{a}=m^{-1}\tilde{\rho}^{a}$, where $m$ is a constant with the dimension of a mass, and define the dimensionless metric the same as in the previous expression,
\begin{equation}
    \label{eq:geometricalmetricdefinition}
    g=\eta_{ab}\rho^{a}\otimes\rho^{b}=m^{-2}\eta_{ab}\tilde{\rho}^{a}\otimes\tilde{\rho}^{b}.
\end{equation}
However, this equation does not actually define a metric on $M$ since, strictly speaking, 
the tensor constructed in this way lies in $T^{*}P\otimes_{\text{sym}}T^{*}P$, where $\otimes_{\text{sym}}$ denotes 
the symmetrized inner product.
We then consider $\rho^{a}$ in Eq.~(\ref{eq:geometricalmetricdefinition}) to be the local connection 1-form (on $M$) associated with $\pmb{\rho}$ and a section $s\in\Gamma(M;P)$ (here $\Gamma(M;P)$ stands for the set of sections of the principal G-bundle $P$).
Notice that making another choice for $s$ changes the metric definition, unless the change of gauge ({\it i.e.} a change of section $s\rightarrow s'$) leaves Eq.~(\ref{eq:geometricalmetricdefinition}) invariant.
Figure~\ref{figure one} shows the original 5-dimensional manifold~\footnote{
This 5-dimensional manifold can be either flat -- in which case it can be considered to be identical to the space over which the fibration is constructed -- or 
it may be curved. The difference between these two cases is not relevant for this work, and therefore 
it will not be discussed any further.} (purple) over which the fibration $P$ is 
constructed, and the spacetime indices
manifold $M$ is denoted by the yellow line of co-dimension one.
The sections $\Gamma(M,P)$ are then defined as the sections of the fibration $P$ intersecting $M$. The original gauge group $G$ 
-- defined by the fibration $P$ of the 5-dimensional manifold --
is broken by the choice of a section $s \in\Gamma(M,P)$ to $SO(1,3)$, which is the symmetry group of general relativity.
Indeed, the only gauge transformations that preserve the gauge theoretical metric tensor are those which act pseudo-orthogonally on the gauge tetrad fields, {\it i.e.}
\begin{equation*}
    \rho^{i}\rightarrow\Matrice{\Lambda}{a}{b}\rho^{b},\qquad \Lambda\in SO(1,3)\subset G.
\end{equation*}
With the construction above we turned $M$ into a pseudo-Riemannian manifold $(M,g)$ with Lorentzian signature.
Now that we established the tetrad nature of part of the gauge field, we will use $\rho^{a}\equiv e^{a}$ in order to adapt to the standard notation.

We introduce the standard action functional of Yang-Mills theory,
\begin{equation}
    \label{eq:actionfunctional}
    S[\pmb{\omega}]=\int_M \langle\Omega,\Omega\rangle
    =\int_{M}\Omega^{i}\wedge*\Omega^{j}G_{ij}=\int_{M}{\rm d}^{4}x\,\text{det}(e)\,\Matrice{\Omega}{i}{\mu\nu}\Omega^{j\mu\nu}G_{ij}\,,\quad
\end{equation}
where $\langle\cdot ,\cdot\rangle$ denotes the inner product
on the group space
and $i=1,2,\cdots, N$ is the algebra index and 
$N={\rm dim}[\mathfrak{g}]$.
It is evident that in general the theory~(\ref{eq:actionfunctional})
 is Lorentz invariant and not $G$-invariant.
This happens because the Hodge-star operator introduces in the action a non-trivial metric dependence
which, as pointed out above, induces a symmetry 
breaking.~\footnote{One can add to the action~(\ref{eq:actionfunctional}) a term
$\propto \int_M\langle\Omega,*\Omega\rangle$, which 
is fully gauge invariant. Such a term is 
({\it via} the Hodge dual) endowed with the spacetime Levi-Civita tensor, and in this way resembles 
the MacDowell-Mansouri and Wilczek theories.
However, in our rendering of the inner product this term is purely 
topological, meaning that it does {\it not} 
contribute to the equations of motion, and thus does not in any 
way affect the theory at the classical level, analysed in this work.
In particular, it cannot give the Hilbert-Einstein action as its low-energy limit.}
We will then give a non-degenerate inner product to $\mathfrak{g}$ that is at least Lorentz invariant, $G_{ij}$.
The equations of motion are obtained by Hamilton's variational principle and read:
\begin{equation}
    \label{eq:geometriceomproto}
    \begin{aligned}
    G_{aj}\!\left[\frac{1}{\sqrt{-g}}\de{\gamma}\left(\sqrt{-g}\Omega^{j\delta\gamma}\right)\!+\!\matrice{c}{lm}{j}\Matrice{\omega}{l}{\gamma}\Omega^{m\delta\gamma}\right]\!
    &=G_{ij}\left[\Omega^{i\delta\nu}\Matrice{\Omega}{j}{\mu\nu}\matrice{e}{a}{\mu}\!-\!\frac{\matrice{e}{a}{\delta}}{4}\Matrice{\Omega}{i}{\mu\nu}\Omega^{j\mu\nu}\right]\bigg|_{\text{if $a\in{\cal S}_e$ 
     }},\quad\\
    G_{aj}\!\left[\frac{1}{\sqrt{-g}}\de{\gamma}\left(\sqrt{-g}\Omega^{j\delta\gamma}\right)
    \!+\!\matrice{c}{lm}{j}\Matrice{\omega}{l}{\gamma}\Omega^{m\delta\gamma}\right]\!&=0\bigg|_{\text{if $a\notin{\cal S}_e$ 
    }}\,,
    \end{aligned}
\end{equation}
where ${\cal S}_e=\{0,1,2,3\}$ denotes the set of tetrad indices.
We found the peculiarity of a geometric Yang-Mills theory. The gauge fields that take the role of the tetrad are not source-free in vacuum, yet they are sourced by the geometric energy-momentum tensor. 
In the following we focus on studying 
pseudo-orthogonal gauge theories, {\it i.e.} $\omega^{i}=\omega^{[AB]}$.


\section{De Sitter Yang-Mills theory}
\label{sec:second}

In this section we study the geometrical Yang-Mills theory for the de Sitter group $SO(1,4)$.
We show that the geometrical Yang-Mills action contains the Hilbert action in the presence of a cosmological constant.
We advice the reader to go through Appendix \ref{appendix:pseudoorthogonalgroups} to get more familiar with the techniques we will be using.
In the following we use lower-case latin letters for the Lie algebra indices corresponding to the Lorentz generators,
{\it i.e.} $M_{AB}|_{A,B=0,...,3}\equiv M_{[ab]}$.
The other four generators, $M_{[a4]}\equiv\tilde{P}_{a}$, generate translations.
Comparing with Eq.~(\ref{eq:pseudoorthogonalcomputingcommutators}),  we get the commutators 
in the Lie algebra basis with this new notation,
\begin{equation}
    \label{eq:desittercommutators}
    \begin{aligned}
        \comm{M_{[ab]}}{M_{[cd]}}&=\eta_{bc}M_{[ad]}+\eta_{ad}M_{[bc]}+\eta_{db}M_{[ca]}+\eta_{ac}M_{[db]},\\
        \comm{M_{[ab]}}{\tilde{P}_{c}}&=\eta_{bc}\tilde{P}_{a}-\eta_{ac}\tilde{P}_{b},\\
        \comm{\tilde{P}_{a}}{\tilde{P}_{c}}&=M_{ca},
    \end{aligned}
\end{equation}
which gives for the Killing metric,
\begin{equation}
    \label{eq:desitterkilling}
    \begin{aligned} 
        G_{[ab][cd]}&=2(D-2)\left[\eta_{bc}\eta_{da}-\eta_{bd}\eta_{ca}\right],\\
        G_{[ab][c4]}&=0,\\
        G_{[a4][c4]}&=-2(D-2)\eta_{ac},
    \end{aligned}
\end{equation}
where here $D=5$
 (compare with Eq.~(\ref{eq:pseudoorthogonalcomputingkilling})).

We now consider the particular pseudo-orthogonal bundle for which the structure group is given by de Sitter group.
As usual we introduce a connection,
\begin{equation}
    \label{eq:desitterconnection}
    \pmb{\omega}=\frac{1}{2}\omega^{[AB]}\otimes M_{[AB]}=\frac{1}{2}\omega^{[ab]}\otimes M_{[ab]}+e^{a}\otimes\tilde{P}_{a},
\end{equation}
which gives for the curvature (compare with Eq.~(\ref{eq:pseudoorthogonalcurvature})):
\begin{equation}
    \label{eq:desittercurvature}
    \begin{aligned}
        \pmb{\Omega}&={\rm d}\pmb{\omega}+\frac{1}{2}\comm{\pmb{\omega}}{\pmb{\omega}}\\
        &=\frac{1}{2}\left[{\rm d}\omega^{[ab]}+\Matrice{\omega}{[a}{c]}\wedge\omega^{[cb]}
                   -e^{a}\wedge e^{b}\right]\otimes M_{ab}
                   + \left[{\rm d}e^{a}+\Matrice{\omega}{[a}{b]}\wedge e^{b}\right]\otimes \tilde{P}_{a}\\
        &\equiv \frac{1}{2}\Omega^{[ab]}\otimes M_{[ab]}+T^{a}\otimes \tilde{P}_{a}
\,.\qquad
    \end{aligned}
\end{equation}
The Bianchi identities are given by,
\begin{equation}
    \label{eq:pseudoorthogonalcovariantderivativeforms}
    \begin{aligned}
    {\rm d}_{\pmb{\omega}}\pmb{\Omega}&={\rm d}\pmb{\Omega}+\comm{\pmb{\omega}}{\pmb{\Omega}}\\
    &=\frac{1}{2}\left[{\rm d}\Omega^{[AB]}+\Matrice{\omega}{[A}{C]}\wedge \Omega^{[CB]}-\Matrice{\omega}{[B}{C]}\wedge \Omega^{[CA]}\right]=0.
    \end{aligned}
\end{equation}

We will use the fields $\Labeledset{e}{a}{a}$ as tetrad fields, in terms of which 
we can define the metric as in Section \ref{sec:first},
\begin{equation}
    \label{eq:desitterdefinitionmetric}
    g=\eta_{ab}e^{a}\otimes e^{b}.
\end{equation}
The other part of the connection 1-form is related to the Lorenz generators and it corresponds 
to the covariant derivative on the tangent bundle of $M$.
Under the interpretation explained above, the curvature related to the tetrad fields is given by
the torsion on $M$ related to the covariant derivative inherited by $\omega^{[ab]}$.
We introduce the notation,
\begin{equation}
    \label{eq;desittersimilriemann}
    R^{[ab]}={\rm d}\omega^{[ab]}+\Matrice{\omega}{[a}{c]}\wedge\omega^{[cb]},
\end{equation}
so that we can write,
\begin{equation}
    \label{eq:desitterrenamingomega}
    \Omega^{[ab]}=R^{[ab]}-\tilde{e}^{a}\wedge \tilde{e}^{b}=R^{[ab]}-m^{2}e^{a}\wedge e^{b}.
\end{equation}
It is well known that a torsionless, metric-compatible
connection (such as $\omega_{[ab]}$)
automatically fixes it to be the Levi-Civita connection.
Indeed, we see that for the configurations for which $T^{a}=0$, we have,
\begin{equation}
    \label{eq:desitterriemannequivalence}
    \overset{\!\!\!\circ}{R^{ab}}=R^{[ab]},
\end{equation}
where $\overset{\!\!\!\circ}{R^{ab}}$ is the Riemann curvature tensor.
This equivalence is extremely important now that we will build the action.

Following the recipe of Section \ref{sec:first} we provide the action for 
the geometrical Yang-Mills theory for the de Sitter group,
\begin{equation}
    \label{eq:desitteraction}
    \begin{aligned}
        S[\omega^{[ab]},e^{a}]\!&=\!\alpha_0\int_{M}\!\frac{1}{4}\Omega^{[ab]}\!\wedge*\Omega^{[cd]}G_{[ab][cd]}
        +{T}^{a}\!\wedge* {T}^{b}G_{[a4][b4]}\\
        &=\alpha\!\int_{M}\!\frac{1}{2}\Matrice{\Omega}{[a}{c]}\wedge*\Matrice{\Omega}{[c}{a]}-m^{2}T^{a}\wedge* T_{a}\\
        &=\alpha\!\int_{M}\!\frac{1}{2}\Matrice{R}{[a}{c]}\wedge* \Matrice{R}{[c}{a]}\!-\!m^{2}T^{a}\wedge* T_{a}
        \!-\!m^{2}e^{a}\!\wedge e^{c}\wedge* R_{ca}\!+\!\frac{m^{4}}{2}e^{a}\wedge e^{c}\wedge*\!\left(e_{c}\wedge e_{a}\right)\\
        &=\alpha\int_{M}\sqrt{-g}\,{\rm d}x^4\bigg[\!-\frac{1}{4}R^{[ac]\mu\nu}R_{[ac]\mu\nu}
        +\frac{m^{2}}{2}T^{[a]\mu\nu}T_{[a]\mu\nu}+m^{2}(R-2\Lambda)\bigg],
    \end{aligned}
\end{equation}
where $\alpha =2(D-2)\alpha_0=6\alpha_0$ as in Eq.~(\ref{eq:pseudoorthogonalcomputingkilling}),
$\alpha_0$ is the (inverse) gauge coupling constant of the original gauge theory
and $\Lambda=\frac{n(n-1)}{4}m^{2}=3m^{2}$ is the 
{\it (gauge theoretical) cosmological constant} coming from the last term in the above equations.
Recall that we introduced the mass parameter $m$ in accordance with dimensional analysis. 
For the theory~(\ref{eq:desitteraction}) is to reduce in the low energy limit to general relativity, from the
 last line of~(\ref{eq:desitteraction})
it follows that $\alpha m^2\rightarrow \frac{1}{16\pi G} = \frac{M_{\rm Pl}^2}{2}$ and 
$\Lambda = 3m^2 =  \frac{3}{16\pi G \alpha} = \frac{3M_{\rm Pl}^2}{2\alpha}$, such that $\alpha$ can 
be used to tune the cosmological constant. In particular, when $\alpha\gg 1$,~\footnote{Recall that in gauge theories
$\alpha$ is related to the gauge coupling constant as, 
$\alpha =1/(2g^2)$, such that $\alpha\gg 1$ corresponds to the weak coupling limit.} the geometrical cosmological constant 
is much smaller than the Planck scale. 
While the action~(\ref{eq:desitteraction}) still exhibits all physical degrees of freedom, its importance is in its torsionless limit,
}
\begin{equation}
    \label{eq:desittertorsionlessaction}
    \begin{aligned}
        S[\omega^{[ab]},e^{a}]&=\alpha\int_{M}\frac{1}{2}\Matrice{\overset{\circ}{R}}{[a}{c]}\wedge* \Matrice{\overset{\circ}{R}}{[c}{a]}-m^{2}e^{a}\wedge e^{c}\wedge* \overset{\circ}{R}_{ca}
         +\frac{m^{4}}{4}e^{a}\wedge e^{c}\wedge*\left(e_{c}\wedge e_{a}\right)\\
        &
        =\alpha\!\int_{M}\sqrt{-g}\,{\rm d}x^4\left[\!-\frac{1}{4}\overset{\circ}{R}^{[ac]\mu\nu}\overset{\circ}{R}_{[ac]\mu\nu}\!+\!m^{2}(\overset{\circ}{R}\!-\!2\Lambda)\right],\;\;
    \end{aligned}
\end{equation} 
which gives, as mentioned above, the Einstein-Hilbert action supplemented by a cosmological constant and a Riemann squared term.
The latter is not multiplied by $m^2$,
and this mass scale needs to be high enough~\footnote{The mass $m$ ought to be higher than 
 the energy scale at which general relativity is well tested.} 
This means that the effect of this interaction on the Hilbert action is suppressed as $1/m^2$.
The equations of motion follow from Eq.~(\ref{eq:geometriceomproto}) and read,
\begin{equation}
    \label{eq:desittereom}
\begin{aligned}
    &G_{[ab][cd]}\bigg[\frac{1}{2\sqrt{-g}}\de{\gamma}\left(\sqrt{-g}\Omega^{[cd]\delta\gamma}\right)
    \!+\!\frac{1}{4}\matrice{c}{[ef][lm]}{\quad\quad [cd]}\Matrice{\omega}{[ef]}{\gamma}\Omega^{[lm]\delta\gamma}
    \!+\!m^{2}\matrice{c}{[4f][4m]}{\quad\quad [cd]}\Matrice{e}{f}{\gamma}T^{m\delta\gamma}\bigg]\!=0,\\
    &\frac{1}{\sqrt{-g}}\de{\gamma}\left(\sqrt{-g}\matrice{T}{a}{\delta\gamma}\right)\!+\!\omega_{[an]\gamma}T^{n\delta\gamma}\!+\!\matrice{\text{Ric}}{a}{\delta}\!-\!\frac{\matrice{e}{a}{\delta}}{2}\left(R\!-\!2\Lambda\right)
    =\frac{1}{2m^{2}}\matrice{\left(\Theta_{\text{lorentz}}\right)}{a}{\delta}\!+\!\matrice{\left(\Theta_{\text{torsion}}\right)}{a}{\delta}
    ,
\end{aligned}
\end{equation}
where we identified,
\begin{equation}
    \label{eq:desittereomtetradinfo}
    \begin{aligned}
        \matrice{\text{Ric}}{a}{\delta}&=R^{[cd]\delta\nu}\eta_{ac}e_{d\nu},\\
        \matrice{\left(\Theta_{\text{lorentz}}\right)}{a}{\delta}&=\matrice{e}{a}{\mu}R^{[cd]\delta\nu}R_{[cd]\mu\nu}-\frac{1}{4}\matrice{e}{a}{\delta}R^{[cd]\mu\nu}R_{[cd]\mu\nu},\\
        \matrice{\left(\Theta_{\text{torsion}}\right)}{a}{\delta}&=\matrice{e}{a}{\mu}T^{d\delta\nu}T_{d\mu\nu}-\frac{1}{4}\matrice{e}{a}{\delta}T^{d\mu\nu}T_{d\mu\nu}.
    \end{aligned}
\end{equation}
Once again, it is interesting to study torsionless solutions to the equations of motion.
We see that they correspond to Einstein's field equations supplemented by a geometrical energy momentum tensor and the corresponding equations for the Lorentz connection, namely:
\begin{equation}
    \label{eq:desittertorsionlesseomtetrad}
    \begin{aligned}
    \matrice{\overset{\circ}{\text{Ric}}}{a}{\delta}-\frac{\matrice{e}{a}{\delta}}{2}\left(\overset{\circ}{R}-2\Lambda\right)&=\,\frac{1}{2m^{2}}\matrice{\left(\Theta_{\text{lorentz}}\right)}{a}{\delta},\\
    G_{[ab][cd]}\bigg[\frac{1}{2\sqrt{-g}}\de{\gamma}\left(\sqrt{-g}\Omega^{[cd]\delta\gamma}\right)
    +\frac{1}{4}\matrice{c}{[ef][lm]}{\quad\quad [cd]}\Matrice{\omega}{[ef]}{\gamma}\Omega^{[lm]\delta\gamma}\bigg]&=0.
    \end{aligned}
\end{equation}
This result is somewhat surprising. 
Indeed, we obtained Einstein's equations and a cosmological constant from the standard Yang-Mills action.
In other words, we derived the equations of the gravitational field from a theory more similar to QCD or the Electro-weak interaction, and in general to the Standard Model physics.
Moreover, notice that the difference between proper GR and Eq.~(\ref{eq:desittertorsionlesseomtetrad}) is a factor which is of second order in the Riemann curvature tensor but is also suppressed by an inverse Planck mass squared.
The contribution coming from a non-vanishing right-hand-side
 of the vacuum Einstein's equation is relevant only when the curvature is of the same order of magnitude as the Planck mass. This is the situation one usually finds 
close to singularities of the GR solutions.  
Adding matter to the theory will result in a contribution on the right-hand-side of 
Eq.~(\ref{eq:desittertorsionlesseomtetrad}).
For the tetrad equation one would find the energy-momentum tensor of the matter field, 
since it would appear from the Hodge-star variation.
For the Lorentz connection one would get the contribution coming from the gauge current (as in standard 
Yang-Mills theory) which would be represented by the angular momentum of the matter fields (since the gauge symmetry group is given by local Lorentz transformations).

An important aspect of this theory is the appearance of a cosmological constant.
This constant is positive and it is proportional to the Planck mass squared, 
{
more precisely $\Lambda\sim m^2\sim M_{\rm Pl}^2/\alpha$,
which is also the curvature scale at which the gravitational energy from curvature becomes dynamically important.
Given that the electroweak scale transition changes the vacuum energy density by an amount $\Delta \rho \sim E_{\rm EW}^4$,
which contributes to the cosmological constant as, $\Delta\Lambda_{EW}\sim E_{\rm EW}^4/M_{\rm Pl}^2$. This then provides an 
upper bound on $m^2\sim M_{\rm Pl}^2/\alpha < E_{\rm EW}^4/M_{\rm Pl}^2$, from which we conclude,
$\alpha = 1/(4g^2) > (M_{\rm Pl}/E_{\rm EW})^4 \sim 10^{64}$, or equivalently $g< 10^{-32}$, a tiny gauge coupling 
constant.~\footnote{If $m\sim E_{GUT}\sim 10^{16}~{\rm GeV}$ were of the order of the grand unified scale,
the constraint on $\alpha$ and $g$ would be much milder,  
$g\sim  (E_{\rm GUT}/M_{\rm Pl})^2\sim 10^{-6}$, which is of the order of the electron yukawa.}
However, these relations holds classically, and they will change when quantum (loop) contributions to the cosmological constant are included. 
To summarize, the scale $m$ is the scale above which the geometric theory of gravity behaves as gauge theory.
Current observations suggest that the scale $m$ is of the order of the electroweak scale (or grand unified scale, if grand unification
was realised). However, due to the smallness of the gravitational gauge coupling constant, this gravitational theory does not affect 
significanly particle physics experiments, and therefore accelerator physics
experiments cannot yet constrain the theory.

Consider again the complete de Sitter geometrical action in Eq.~(\ref{eq:desitteraction}).
We would like to give an intuitive scheme that one could follow in order to constrain dynamically the second term (torsion squared) to be zero.
Let $A=A_{\mu}{\rm d}x^{\mu}=A_{a}e^{a}$ be a 1-form vector field on the spacetime $M$, not necessarily a gauge boson. 
Here we exploited the interpretation of part the connection field as tetrad fields to express the coordinates of the $A$ field in this orthonormal basis.
Recalling that the tetrad generators correspond to $M_{[a4]}$ one can see from 
Eq.~(\ref{eq:pseudoorthogonalcovariantderivativeforms}) that the covariant derivative acting on the tetrad fields is given by,
\begin{equation}
    \label{eq:desittercovariantderivativetetrad}
    {\rm d}_{\pmb{\omega}}e^{a}={\rm d}e^{a}+\Matrice{\omega}{[a}{b]}e^{b}=T^{a},
\end{equation}
so that charging the covector fields $A$ as a Lorentz multiplet~\footnote{This means that we are applying the principle of general covariance passing from rigid to local Lorentz transformation
acting on the covector. We need Lorentz and not the entire general linear group, as would be the case for general coordinate invariance, since we can always cover a Lorentzian manifold 
with local orthonormal frames for the tangent bundle.}, we find:
\begin{equation}
    \label{eq:desittercovariantderivativecovector}
    {\rm d}_{\pmb{\omega}}A={\rm d}A_{a}\wedge e^{a}+A_{a}T^{a}.
\end{equation}
It clearly provides a gauge equivariant expression and it is different from the standard exterior derivative only 
if torsion is non-vanishing (as it is the case when one consider the general covariance principle applied to electro-magnetism).
The easiest term to include in the action for such a field would be the standard kinetic term,
\begin{eqnarray}
    \label{eq:desitterkineticcovector}
    \int_{M}F\wedge* F&\!\equiv\!\!&\int_{M}{\rm d}_{\pmb{\omega}}A\wedge*{\rm d}_{\pmb{\omega}}A
    \\
    &\!\!=\!\!&\int_{M}\bigg[{\rm d}A_{a}\wedge e^{a}\wedge*\left({\rm d}A_{b}\wedge e^{b}\right)
    \!+\!2A_{a}T^{a}\wedge*\left({\rm d}A_{b}\wedge e^{b}\right)\!+\!A_{a}A_{b} T^{a}\wedge *T^{b}\bigg].
 \nonumber
\end{eqnarray}
We now see that, if one introduces a suitable potential for the $A$-field, there is the possibility of a condensation of the field such to give $\left\langle A_{a}A_{b}\right\rangle=-\frac{1}{2}m^{2}\eta_{ab}$.
The semiclassical limit would then correspond to the torsionless action in Eq.~(\ref{eq:desittertorsionlessaction}), turning the de Sitter gauge theory into an Einstein-Cartan theory~\cite{Cartan:1923zea,Hehl:1976kj}. 
Since the only dynamical massless vector field in the Universe is the photon, 
this gravitational vector field should be massive enough not to be detectable by modern experiments.
Furthermore, since there is no dependence in the action on the derivative of the tetrad fields, 
 the torsion field becomes non-dynamical and once we fix the initial conditions to give a vanishing torsion this 
will remain true throughout evolution. 

In conclusion, using the de Sitter group as gauge group for a geometrical Yang-Mills theory we are able to obtain Einstein's theory of gravity as a low energy torsionless limit of our theory.
The Yang-Mills formulation, and in particular the structure constants of the de Sitter algebra, give the Hilbert action supplemented with a Riemann squared term, 
which is suppressed by a Planck mass squared, a torsion squared factor, which could be possibly removed dynamically as we have shown in Eq.~(\ref{eq:desitterkineticcovector}),
and a cosmological costant (as well as the usual mass parameter expected in all geometrical Yang-Mils theories).
In Section~\ref{sec:third} we 
perform a Hamiltonian analysis suitable for geometrical Yang-Mills theories.
In particular we consider the constraints arising in phase space and we provide their analysis.

It is worth noticing that, upon replacing de Sitter group SO(1,4) with anti-de Sitter group SO(2,3), one would find the same results we have found for de Sitter since the algebra of the two groups is very similar,
in particular the theory would still contain the Einstein-Hilbert action.
The difference lies in the cosmological constant, which would be negative for the case of AdS gauge theory. This comes from some relative sign between the structure constants of $\mathfrak{so}(1,4)$ and $\mathfrak{so}(2,3)$.


\section{Hamiltonian analysis}
\label{sec:third}
\label{Hamiltonian analysis}

In order to specify a Hamiltonian~\cite{Prokhorov:2011zz} for the theories at hand, we need to break covariance with respect to coordinates specifying a time (read evolution) direction.
As usual, the Yang-Mills Lagrangian is {\it singular}, {\it i.e.} its Hessian is degenerate: $\text{det}\left(T^{ij}\right)=\text{det}\left(\delta^2L/\delta\dot{\omega}_{i}\delta\dot{\omega}_{j}\right)=0$.
This implies that constraints will arise in phase space.
In the following non-tetrad indices will be given by $[AB]$ while for tetrad fields we will use $[\cdot a]$.
The extended Hamiltonian is given by,
\begin{eqnarray}
    \label{eq:totalhamiltonian}
        H_{T}\!\!&=&\!\!\int_{\rn{3}}d^{3}x\bigg\{\!-\frac{1}{4\alpha\sqrt{-g}}\matrice{\Pi}{[CD]}{l}\Pi^{[CD]k}\inversem{lk}
           \!+\!\frac{1}{2\alpha\sqrt{-g}}\matrice{\Pi}{[CD]}{l}P^{[CD]k}\inversem{lk}
\nonumber\\
        \!\!&&\!\hskip 1.5cm
        \!-\,\alpha\sqrt{-g}\left(g^{k0}g^{ij}\Omega_{[AB]ki}\right)\left(g^{s0}g^{tl}\Matrice{\Omega}{[AB]}{st}\right)\inversem{lj}\;\;
     \!+\!\frac12\alpha\sqrt{-g}g^{ij}g^{kl}\Omega_{[AB]jk}\Matrice{\Omega}{[AB]}{il}
\nonumber\\
        \!\!&&\!\hskip 1.5cm
        \!+\,u^{[CD]}\phi_{[CD]}\bigg\},
\label{extended hamiltonian}
\end{eqnarray}
where the last term stands for all the Lagrange multipliers and constraints that arise during the analysis of the primary constraints $\phi_{[CD]}=\matrice{\Pi}{[CD]}{0}\approx 0$,
where $\approx$ stands for weak ({\it on-shell}\/) equality.
These are the standard primary constraints of Yang-Mills theories.
The secondary constraints are given by the usual {\it generalized} Gauss' law.
For non-tetrad fields we have,
\begin{equation}
    \label{eq:nontetradsecondaryconstraints}
    \begin{aligned}
    0&\approx -\del{H_{T}}{\Matrice{\omega}{[CD]}{0}}=D_{i}\matrice{\Pi}{[CD]}{i}\equiv\\
    &\equiv\left[\de{i}\matrice{\Pi}{[CD]}{i}+\matrice{\Pi}{[CA]}{i}\matrice{\omega}{[D\,\,\, i}{\,\,A]}-\Matrice{\omega}{[A}{\,\,C]i}\matrice{\Pi}{[AD]}{i}\right],
    \end{aligned}
\end{equation}
while for tetrad fields we find,
\begin{equation}
    \label{eq:tetradsecondaryconstraints}
    \begin{aligned}
    0\approx& \del{H_{T}}{\Matrice{e}{a}{0}}=\bigg\{\!-D_{i}\matrice{\Pi}{(\cdot a)}{i}+\bigg[\Matrice{\Omega}{(CD)}{ki}\matrice{\Pi}{(CD)}{i}
    +2\alpha\sqrt{-g}\,\Omega_{(AB)kl}\left(g^{s0}g^{tl}\Matrice{\Omega}{(AB)}{st}\right)\bigg]\matrice{e}{a}{k}
    \\
    &\hskip 1.5cm
    -\frac12\alpha\sqrt{-g}\bigg[\left(g^{i0}\matrice{e}{a}{j}+g^{j0}\matrice{e}{a}{i}\right)g^{kl}
    \left(g^{k0}\matrice{e}{a}{l}+g^{l0}\matrice{e}{a}{k}\right)g^{ij}\bigg]\Omega_{[AB]jk}\Matrice{\Omega}{[AB]}{il}\\
    &\hskip 1.2cm
    +\matrice{e}{a}{0}\bigg[\!-\frac{1}{4\alpha\sqrt{-g}}\matrice{\Pi}{[CD]}{l}\Pi^{[CD]j}\inversem{lj}
    \!-\!\alpha\sqrt{-g}\left(g^{k0}g^{ij}\Omega_{[AB]ki}\right)\left(g^{s0}g^{tl}\Matrice{\Omega}{[AB]}{st}\right)\inversem{lj}\\
    &\hskip 1.cm
    +\frac12\alpha\sqrt{-g}g^{ij}g^{kl}\Omega_{[AB]jk}\Matrice{\Omega}{[AB]}{il}-\Matrice{\Omega}{[CD]}{ki}\matrice{\Pi}{[CD]}{i}\frac{g^{k0}}{g^{00}}\bigg]\bigg\}.
    \end{aligned}
\end{equation}
These constraints do not generate any new one, yet they impose restrictions on the Lagrange multipliers ($u_{[CD]}$) of the theory in the following sense:
\begin{itemize}
    \item The equation for $u^{(2)[ab]}$ fixes the functions $u^{(2)[\cdot a]}$;
    \item The equation for $u^{(2)[\cdot a]}$ fixes the functions $u^{(1)[\cdot b]}$;
\end{itemize}
which shows that our constraints are either {\it first} or {\it second} class.
The only arbitrary Lagrange multipliers we are left with are the ones corresponding to the Lorentz group.
Once again we find that the gauge symmetry of the theory, and in particular of its phase space, is given in general by the Lorentz subgroup.

Now that we completed the constraints analysis of the theory, we can compute the equations of motion using the standard Poisson brackets, namely,
\begin{equation}
    \label{eq:generalizedhamiltonequationsbrackets}
    \Matrice{\dot{\omega}}{[CD]}{\mu}=\Big\{\Matrice{\omega}{[CD]}{\mu},H_{T}\Big\}
    \,,\qquad 
    \matrice{\dot{\Pi}}{[CD]}{\mu}=\Big\{\matrice{\Pi}{[CD]}{\mu},H_{T}\Big\}
    \,.\quad
\end{equation}
One can show that these equations
are equivalent to Eqs.~(\ref{eq:geometriceomproto}), in particular they are covariant.

From the Hamiltonian one can already see a potential problem of our theory.
Notice that we can rewrite the first two terms in Eq.~(\ref{eq:totalhamiltonian}) as,
\begin{equation}
    \label{eq:kineticenergy}
    \begin{aligned}
    \int_{\rn{3}}&\frac{1}{2\alpha\sqrt{-g}}\inversem{lk}\Big[\!-\Big(\matrice{\Pi}{[CD]}{l}\!-\!\matrice{P}{[CD]}{l}\Big)\Big(\Pi^{[CD]k}\!-\!P^{[CD]k}\Big)
    +\matrice{P}{[CD]}{l}P^{[CD]k}\Big]
    \,,
    \end{aligned}
\end{equation}
which is the kinetic energy (first term) and the leftover from completing the square (second term).
Since for pseudo-orthogonal groups the metric in Lie algebra space we use is often an indefinite inner product,
there are {\it kinetic instabilities} in the theory, {\it i.e.} fields for which the kinetic energy comes with the 
negative sign in the total hamiltonian.
Notice that there is no indefiniteness coming from the inner product in spacetime indices.
This is due to the presence of the primary constraints that force the timelike component of the momenta to vanish.
In the constraints analysis we gave we did not find any constraint able to render 
the kinetic instabilities unphysical.
However, the presence of second-class constraints suggests that the simplectic structure of phase space is not canonical and thus not all hope is gone that after one properly identifies the physical
phase space of the theory there will not be this kind of issues.
Another possible solution is to consider the Born-Infeld theory of electromagnetism~\cite{Born:1934gh} replacing the action in Eq.~(\ref{eq:actionfunctional}) with:
\begin{equation}
    \label{eq:borninfeldaction}
    S=2\alpha\beta^{2}\int_{M}{\rm d}^{4}x\,\text{det}(e)
    \left[\sqrt{1\!+\!\frac{1}{\beta^{2}}\Matrice{\Omega}{i}{\mu\nu}\Omega^{j\mu\nu}G_{ij}}
    \!-\!1\right].
\end{equation}
In this way one can covariantly introduce higher powers of the momenta into the Hamiltonian in order to make the kinetic energy bounded from below, thus stabilizing the theory, at least at the classical level. 
This theory harbors higher order interactions, and therefore we leave the Hamiltonian analysis for 
a future work.


\section{Conclusion and outlook}
\label{Conclusion and outlook}

The main goal of this paper was to reformulate general relativity in the context of Yang-Mills theories.
Indeed, it is well-known that Einstein's theory suffers from singularity problems~\cite{Misner:1973prb}
 and it was proved that the theory is not renormalizable~(see~\cite{Buchbinder:2021wzv} for a review).
Knowing that gravity can be formulated as an effective field theory~\cite{Donoghue:2012zc}, we showed that it is possible
to formulate general relativity
as a constrained version of some more general theory in its low energy limit.

Throughout this paper we have shown the intimate relation between pseudo-orthogonal Yang-Mills gauge fields and gravitational theories.
In order to introduce a gauge theoretical cotetrad fields, we have developed a new class of theories, {\it geometrical Yang-Mills theories}~\cite{Hamilton:2017gbn}.
Defining the metric through these particular gauge fields requires the introduction of a mass parameter which will take the role of a
Planck mass.
The geometrical action one obtains is invariant only with respect to the Lorentz subgroup.
The equations of motion one obtains~(\ref{eq:geometriceomproto}) 
are the standard curved spacetime generalization of the well-known Yang-Mills equations.
In the case of the tetrad fields, one sees that these fields are sourced in the vacuum by an energy-momentum tensor related to the field strength of our gauge connection.

The main result of the paper is the de Sitter gauge theory we developed in Section~\ref{sec:second}.
The torsionless low energy limit of the theory coincides with general relativity with the appearance of a positive cosmological constant,
{ whose size is controlled by the Planck scale and the gauge coupling constant.
In the weak coupling limit this geometrical cosmological constant can be much smaller than the Planck scale,
and can be (in part) cancelled by the contributions from quantum fields, to yield the observed cosmological constant. 
These results allow us to consider Einstein's theory of gravity as part of a more general Lorentz invariant de Sitter Yang-Mills theory.
 We argue that the theory might be renormalizable since the vertex structure of the theory is essentially the same as ordinary flat-space Yang-Mills theories.
However, since we studied arbitrary curved spacetimes, there will be new counterterms associated with the non-trivial geometry of spacetime
and the issue of renormalizability of the full quantum theory still requires a careful investigation. 
In the context of quantum field theory in curved spacetime (with a classical gravitational field), it is known that higher-order geometric scalars are induced by quantum corrections. 
These counterterms can be added to the theory using our geometrical Yang-Mills formalism,
avoiding Ostrogradsky instabilities, as we have shown in Section~\ref{sec:third}.
Indeed, the phase space we identified is not the same as for Palatini gravity~\cite{Palatini:1919}
 since the canonical momentum of the metric is not given by the spin connection.
 From the analysis in Section~\ref{Hamiltonian analysis} it follows that the number of phase-space
 degrees of freedom in a gravitational theory
 which originates from a gauge theory (in its torsionless limit) is twice as large as that in the 
Palatini formulation.
These results inspire us to say that geometrical Yang-Mills theories could be a useful formulation of gravitational theories in general.

We provided the first steps towards canonical quantization ($\comm{}{}=i\hbar \{,\}$) building the Hamiltonian and studying the phase space of the theory.
Since pseudo-orthogonal groups are non-compact Lie groups, the Killing form we use in Lie algebra space gives rise to an indefinite inner product.
This means that some fields will pick the wrong sign in the kinetic energy and they could generate kinetic instabilities
in the theory.
However, as one can see from the kinetic energy of the theory, there is no sign of Ostrogradsky instabilities in the theory at hand, as we originally expected, since in Yang-Mills theories the Riemann squared term
is part of the gauge theory.
As it is usually the case with gauge theories, we find both primary and secondary constraints.
There are both first and second-class constraints and their self-consistency conditions reduce the gauge redundancy in phase space to its Lorentz subgroup of the original gauge group.
There are no new constraints arising in phase space and thus the Hamiltonian we provide is complete.
As we argue in Section~\ref{Geometrical Yang-Mills theories}, 
the projection onto the spacetime manifold $M$ of the gauge theory induces the metric, thus breaking the gauge symmetry to its Lorentz subgroup, the symmetry group of general relativity (see figure~\ref{figure one}).

In conclusion, we provided a new formalism for which interesting results can be found in the context of gravitational theories.
We have showed another way of deriving general relativity out of the geometrical gauge  theory
 and we have provided a consistent Hamiltonian framework suitable for any Yang-Mills theory in dynamical curved spacetime.

\section*{Acknowledgements}
\label{Acknowledgements}

The authors thank Enis Belgacem and Antonino Marciano for numerous discussions and suggestions
that led to significant improvements of the manuscript.
This work is part of the Delta ITP consortium, a program of the Netherlands Organisation
for Scientific Research (NWO) that is funded by the Dutch Ministry of Education, Culture
and Science (OCW) --- NWO project number 24.001.027.


\appendix
\section{Pseudo-orthogonal groups and bundles}
\label{appendix:pseudoorthogonalgroups}
\subsection{Pseudo-orthogonal groups}

In this section we  define pseudo-orthogonal groups and study their properties.
Consider a vector space $\rn{D}$ equipped with the following metric (in the canonical cartesian basis),
\begin{equation}
    \label{eq:pseudoorthogonalmetric}
    \eta=\text{diag}(\underset{S}{\underbrace{-1,..,-1}},\underset{T}{\underbrace{+1,...,+1}}),
\end{equation}
with $S+T=D$.
This inner product turns $\rn{D}$ into the (pseudo-)normed vector space 
$\rn{S,T}$ (for $S=1$ and $T=3$ we get Minkowski spacetime).
We define the (fundamental representation of the) pseudo-orthogonal group 
$O(S,T)$ as the set of transformations on $\rn{S,T}$ that leave the inner product $\eta(X,Y)$ invariant, $X,Y\in\rn{S,T}$,
{\it i.e.} $\eta(\Lambda \cdot X,\Lambda\cdot Y)=\eta(X,Y)$, $\Lambda\in O(S,T)$.
It can be proved that pseudo-orthogonal groups are Lie groups.
It seems then natural to look at transformations infinitesimally close to the identity in order to identify their generators, {\it i.e.} $\Matrice{\Lambda}{a}{b}=\Matrice{\delta}{a}{b}+\epsilon\Matrice{M}{a}{b}+O(\epsilon^{2})$.
We get,
\begin{equation}
    \label{eq:pseudoorthogonalinfinitesimaltransformation}
    \begin{aligned}
    \eta(X,Y)&=\eta_{AB}X^{A}Y^{B}\rightarrow \eta_{AB}X^{A}Y^{B}\\
    &+\epsilon\left[\eta_{AB}\Matrice{M}{A}{C}X^{C}Y^{B}+\eta_{AB}X^{A}\Matrice{M}{B}{C}Y^{C}\right]+O(\epsilon^{2})\\
    &=\eta_{AB}X^{A}Y^{B}+\epsilon X^{C}Y^{B}\left[M_{BC}+M_{CB}\right]+O(\epsilon^{2}),\\
    &\Rightarrow M_{BC}=-M_{CB}
    \,.
    \end{aligned}
\end{equation}
Here and throughout this appendix capital latin indices run from $-S+1,...,0,...,T$.
The previous result shows that the generators of the pseudo-orthogonal group $O(S,T)$ are given by $D\times D$ antisymmetric matrices (when one index is lowered as in Eq.~(\ref{eq:pseudoorthogonalinfinitesimaltransformation})).
There are then ${D(D-1)}/{2}$ linearly independent generators which are given by,
\begin{equation}
    \label{eq:pseudoorthogonalgenerators}
    \Matrice{\left(M_{AB}\right)}{I}{J}=\Matrice{\delta}{I}{A}\eta_{BJ}-\Matrice{\delta}{I}{B}\eta_{AJ}.
\end{equation} 
Notice that $M_{AB}=-M_{BA}$ so that from now on we 
 write $M_{[AB]}$ for the pseudo-orthogonal generators.
In the following we consider only the proper-orthochronous pseudo-orthogonal group ({\it i.e.} the part of $O(S,T)$ which is connected to the identity), so that with the exponential map 
we can recover the whole group.

Now we are ready to study the commutators between the elements of the pseudo-orthogonal Lie algebra.
We compute,
\begin{equation}
    \label{eq:pseudoorthogonalcomputingcommutators}
    \begin{aligned}
        \Matrice{\left(\comm{M_{[AB]}}{M_{[CD]}}\right)}{I}{\!K}=&\Matrice{\left(M_{[AB]}\right)}{I}{\!J}\Matrice{\left(M_{[CD]}\right)}{J}{K} 
        \!-\!\Matrice{\left(M_{[CD]}\right)}{\!I}{\!\!J}\Matrice{\left(M_{[AB]}\right)}{\!J}{\!K}\\
        =&\Matrice{\left(M_{[AD]}\right)}{\!I}{\!\!K}\eta_{BC}\!-\!\Matrice{\left(M_{[AC]}\right)}{\!I}{\!\!K}\eta_{BD}
        \!-\!\Matrice{\left(M_{[BD]}\right)}{\!I}{\!\!K}\eta_{AC}\!+\!\Matrice{\left(M_{[BC]}\right)}{\!I}{\!\!K}\eta_{AD}
        \\
        =&\Big[\Matrice{\Delta}{[EF]}{[AD]}\eta_{BC}-\Matrice{\Delta}{[EF]}{[AC]}\eta_{BD}
        -\Matrice{\Delta}{[EF]}{[BD]}\eta_{AC}+\Matrice{\Delta}{[EF]}{[BC]}\eta_{AD}\Big]\Matrice{\left(M_{EF}\right)}{I}{\!K}\\
        \equiv&\matrice{c}{[AB][CD]}{\quad\quad[EF]}\Matrice{\left(M_{[EF]}\right)}{I}{\!K},
    \end{aligned}
\end{equation}
where we introduced the identity in antisymmetric $\binom{0}{2}$ tensor space, {\it i.e.}
\begin{equation}
    \label{eq:identityantisymmetric}
    \Matrice{\Delta}{[AB]}{[CD]}=\frac{1}{2}\left(\Matrice{\delta}{A}{C}\Matrice{\delta}{B}{D}-\Matrice{\delta}{A}{D}\Matrice{\delta}{B}{C}\right).
\end{equation}
Having identified the structure constants of the pseudo-orthogonal algebra we can now compute the Killing metric for this Lie algebra,
\begin{equation}
    \label{eq:pseudoorthogonalcomputingkilling}
    \begin{aligned}
        G_{[AB][CD]}=&\matrice{c}{[AB][LM]}{\quad\quad[EF]}\matrice{c}{[CD][EF]}{\quad\quad[LM]}\\
        =&\bigg[\eta_{BL}\Matrice{\Delta}{[EF]}{[AM]}+\eta_{AM}\Matrice{\Delta}{[EF]}{[BL]}
        +\eta_{BM}\Matrice{\Delta}{[EF]}{[LA]}+\eta_{AL}\Matrice{\Delta}{[EF]}{[MB]}\bigg]\\
        &\times\!\bigg[\eta_{DE}\Matrice{\Delta}{[LM]}{[CF]}+\eta_{CF}\Matrice{\Delta}{[LM]}{[DE]}
        +\eta_{DF}\Matrice{\Delta}{[LM]}{[EC]}+\eta_{CE}\Matrice{\Delta}{[LM]}{[FD]}\bigg]\\
        =&\,2(D\!-\!2)\left[\eta_{BC}\eta_{DA}\!-\!\eta_{BD}\eta_{CA}\right]\\
        \equiv&\,\alpha\left[\eta_{BC}\eta_{DA}\!-\!\eta_{BD}\eta_{CA}\right]\,.
    \end{aligned}
\end{equation}

\subsection{Pseudo-orthogonal bundles}

Throughout this section we  fix a manifold $M$ and a principal $SO(S,T)$-bundle $P\rightarrow M$.
We introduce a connection 1-form $\pmb{\omega}$ on $P$ and we expand it in the basis of $\mathfrak{so}(S,T)$ in Eq.~(\ref{eq:pseudoorthogonalgenerators}).
\begin{equation}
    \label{eq:pseudoorthogonalconnection}
    \pmb{\omega}=\frac{1}{2}\omega^{[AB]}\otimes M_{[AB]},
\end{equation}
where the $1/2$ in front is necessary to avoid overcounting.
We consider the adjoint bundle $\text{Ad}(P)$.
In particular, we consider the commutator between twisted differential forms ($\pmb{A}=\frac{1}{2}A^{[AB]}\otimes M_{AB},\,A^{[AB]}\in\Omega^{k}(P)$).
We find,
\begin{equation}
    \label{eq:pseudoorthogonalformscommutator}
    \begin{aligned}
        \comm{\pmb{A}}{\pmb{B}}=&\frac{1}{4}A^{[AB]}\wedge B^{[CD]}\otimes\comm{M_{[AB]}}{M_{[CD]}}\\
        =&\frac{1}{4}A^{[AB]}\wedge B^{[CD]}\otimes\bigg(\eta_{BC}M_{[AD]}
        \!+\!\eta_{AD}M_{[BC]}\!+\!\eta_{DB}M_{CA}\!+\!\eta_{AC}M_{DB}\bigg)\\
        =&\frac{1}{2}\left(\Matrice{A}{[A}{C]}\wedge B^{[CB]}-\Matrice{A}{[B}{C]}\wedge B^{[CA]}\right)\otimes M_{[AB]}.
    \end{aligned}
\end{equation}
The curvature associated with $\pmb{\omega}$ is then,
\begin{equation}
    \label{eq:pseudoorthogonalcurvature}
    \begin{aligned}
    \pmb{\Omega}&={\rm d}\pmb{\omega}+\frac{1}{2}\comm{\pmb{\omega}}{\pmb{\omega}}\\
    &=\frac{1}{2}\left[{\rm d}\omega^{[AB]}+\Matrice{\omega}{[A}{C]}\wedge \omega^{[CB]}\right]\otimes M_{[AB]}
    \,,
    \end{aligned}
\end{equation}
and the covariant derivative on twisted forms is given by,
\begin{equation}
    \label{eq:pseudoorthogonalcovariantderivativeforms}
    \begin{aligned}
    {\rm d}_{\pmb{\omega}}\pmb{A}&={\rm d}\pmb{A}+\comm{\pmb{\omega}}{\pmb{A}}\\
    &=\frac{1}{2}\left[{\rm d}A^{[AB]}+\Matrice{\omega}{[A}{C]}\wedge A^{[CB]}-\Matrice{\omega}{[B}{C]}\wedge A^{[CA]}\right]
    .
    \end{aligned}
\end{equation}


\end{document}